\documentclass[apj,a4,onecolumn]{emulateapj}

\begin{document}

\author{Gudlaugur J\'ohannesson, Gunnlaugur Bj\"ornsson and Einar H. Gudmundsson\altaffilmark{*}}
\title{Energy Injection in GRB Afterglow Models}

\altaffiltext{*}{Science Institute, University
  of Iceland, Dunhaga~3, IS--107 Reykjavik, Iceland, e-mail:gudlaugu,
  gulli, einar@raunvis.hi.is}

\begin{abstract}
  We extend the standard fireball model, widely used to interpret 
  gamma-ray burst (GRB) afterglow light curves, to include energy 
  injections, and apply the model to the afterglow light curves of 
  GRB~990510, GRB~000301C and GRB~010222. We show that discrete 
  energy injections can cause temporal variations in the optical 
  light curves and present fits to the light curves of GRB~000301C 
  as an example. A continuous injection may be required to interpret 
  other bursts such as GRB~010222. The extended model accounts 
  reasonably well for the observations in all bands ranging from 
  $X$-rays to radio wavelengths. In some cases, the radio 
  light curves indicate that additional model ingredients may be needed. 
\end{abstract}


\section{Introduction}
After a quarter of a century of modest progress, a breakthrough in
gamma-ray burst (GRB) research came in 1997, when $X$-ray and optical
afterglow emission was first detected from GRB~970228
\citep{vanPar1997,costa1997}, showing a redshift of $z=0.695$.  
This confirmed the cosmological origin of
GRBs and to date, cosmological redshifts have been obtained for about 50 
afterglows\footnote{See {\tt http://www.mpe.mpg.de/$\sim$jcg/grbgen.html}
for the most recent list.}.

Initially, afterglow light curves were often sparsely sampled and 
had a natural explanation in the standard fireball model 
\citep[see e.g.][for a recent review]{Piran2005}.  Many showed a 
smooth power-law decay followed by a steepening one to two days after 
the burst. This temporal behavior has been interpreted as being due to 
a relativistically expanding collimated outflow \citep[e.g.][hereafter
R99]{Rhoads1999} or more recently to a structured jet viewed off axis
\citep[e.g.][]{ZhangMesz2002, Rossi2002}.

The standard fireball model assumes a single energy release event with
subsequent interaction with an ambient medium with constant density or
a stellar wind density profile. It was also generally assumed that the
electron energy distribution index, $p$, was independent of time and 
had values between 2-3, based on theoretical arguments and energy 
considerations \citep[e.g.][]{ZhangMesz2004}.

It is quite remarkable that the light curves of most 
observed afterglows where sufficiently detailed data coverage exists,
can be explained within the standard fireball model or slight
improvements thereof \citep[e.g.][]{Panaitescu2005, Zeh2005}. This most likely 
reflects the underlying universality of the afterglow generation mechanism. 
For some bursts, however, interpretations with the unmodified standard 
model, have proved difficult due to bumps in the light curves or because
some basic model parameters turn out to have values outside the range assumed 
in the model. In particular, the slope of the power law energy distribution 
of the radiating electrons, $p$, is found to be less than 2 in some cases 
\citep[e.g.][]{PanaitescuKumar2001}. \cite{DaiCheng2001} and \cite{Bhatta2001} 
have explored the light curves and spectral signatures in models with $p<2$. 
Numerical calculations of relativistic shock acceleration suggest a universal 
value of $p\approx 2.2-2.3$ \citep[e.g.][]{Achterberg2001}. Other work suggests 
an even higher value of $p\approx 2.6-2.7$ \citep{LemRev2005}. This has 
been referred to as the $p$-problem \citep{Bjornsson2002}. Recent afterglow 
observations do indicate, however, that the value of $p$ may not be 
universal \citep{Shen2005}.

The $p$-problem arises when the standard model is being 
used to interpret light curves that decay much slower than suggested 
by the model. Such a slow decay had in fact been anticipated
\citep[][hereafter RM98 and SM00, respectively]{ReesMesz1998,
  SariMesz2000} by the natural assumption that the mass ejected in
each event (and the associated energy) is released with a range of
Lorentz-factors rather than a single value.  The distributed energy
injection into the fireball sustains higher fluxes for longer periods
of time, thus resulting in slowly decaying light curves, without
requiring $p<2$. The cause of the slow decay is thus shifted from the
electron population to the properties of the source itself.  This
extension of the standard model was, however, not used to interpret
actual data until later \citep[e.g.][]{Nakar2003, Bjornsson2002,
  Bjornsson2004}. As continuous energy injection can sustain slower
light curve decay rates with $p>2$, it is important to explore
fully the properties and light curve signatures of such models. 

In addition to the above quoted references, energy injection in the 
context of GRBs has previously been considered by e.g.\ \cite{PanMeszRees1998} 
who give an example of how the afterglow of GRB~970508 may be explained 
by refreshed shocks, although a spherically symmetric model may also
work. \cite{CohenPiran1999} explore the self-similarity of 
spherical blast waves with continuous energy injection, while 
\cite{DaiLu2000, DaiLu2001} consider a model in which a pulsar is 
assumed to provide the additional energy, a variant of which
was also considered by \cite{ZhangMesz2001}. \cite{KumarPiran2000} explore
a toy model of the interaction of two colliding shells and estimate the 
flux contribution from both the forward and the reverse shocks. 
Then, \cite{ZhangMesz2002} 
consider the interaction of two colliding shells in detail and 
determine a criteria for such collisions to be mild or violent. 
\cite{Panaitescu2005} considers additional energy injection in 
interpreting observations of several GRBs.  
In addition, \cite{GraNaPir2003} and \cite{Huang2005} conclude that
discrete energy injections are needed to properly account for the 
optical afterglow of GRB~030329. All adopt different degrees of 
approximations on various model ingredients.

In recent months, observations by the {\em Swift} satellite have shown
that the early light curve behavior may not be as smooth as previously
thought. The origin of these variations may be due to the transition
from the prompt phase to the afterglow phase or due to a strong
initial reverse shock \citep{PMGBN2005}.  The slower decaying
portions of the light curves may be due to continuous energy injection
\citep{Nousek2005, GranKum2005}.
In particular, GRB~050319 \citep{Cusumano2005} and
GRB~050315, have shown this kind of a behavior 
\citep[see][for the latter burst and further examples]{Chincarini2005}.

In this paper we present a detailed analysis of relativistic fireballs
with either discrete or continuous energy injection. We show that this
has interesting observational consequences in addition to the slower
decay and provides clear signatures in the light curves. 
An example of these effects has already been presented by \cite{Bjornsson2004},
where a discussion of the polarization properties can also be found.  
The clearest of those signatures are bumps in the light curves due to 
refreshed shocks, which may in fact have been observed in e.g.\ GRB~021004 
and GRB~030329. These signatures provide additional information on the 
hydrodynamic properties of the fireball. We give examples of fits to 
three GRB afterglows and show that even with the additional energy 
injection, there remains some discrepancy between the model and the data, 
especially at radio wavelengths. In our approach we try to keep 
the treatment as self-consistent as possible and improve on several 
of the approximations of some of the previous work.

The paper is structured as follows: In Section \ref{sec:dynamics} we
review the effects that energy injections have on the dynamics of the
fireball and outline our approach to the problem. In section
\ref{sec:approx} we present approximations for simple modeling of the
dynamical evolution.  In Section \ref{sec:lcurves} we discuss the
ingredients in the radiative component of our model and 
in Section \ref{sec:interp} we give examples of bursts that
may be interpreted with this modified model.  Finally, in Section
\ref{sec:conc} we discuss our results and conclude the paper.

\section{The Dynamics}
\label{sec:dynamics}
We begin this section by considering energy and momentum
conservation. We follow the approach of RM98 and SM00 and assume that
energy is released over a period that is short
compared to the duration of the afterglow. We then specialize the
method to three cases: An instantaneous energy injection in Section~\ref{subsec:single}
for easy comparison with R99 \citep[see also][]{PaczRhoads93}, 
several discrete injection events in Section~\ref{subsec:discrete}, 
and finally continuous injection with a power law distribution of the Lorentz factors 
in Section~\ref{subsec:continuous}. We assume the 
dynamical evolution to be adiabatic in all cases. 

Assume that the total ejected mass with initial Lorentz factor higher than 
$\Gamma$ can be written as $M(\geq\Gamma)=M_0+\Delta M$, where $M_0$
is the mass ejected with the highest Lorentz factor $\Gamma_0$, and $\Delta
M=\int^\Gamma_{\Gamma_0}(dM/d\Gamma) d\Gamma$, where $dM/d\Gamma$ is a 
function specifying the distribution of ejected mass with Lorentz factor $\Gamma$.
We prefer to use $x=\Gamma/\Gamma_0$ as a variable, and introduce a
dimensionless function $F(x)$, defined by $F(x)=M(\geq\Gamma)/M_0$, in
the interval $1\geq x\geq \Gamma_m/\Gamma_0$, where $\Gamma_m$ is the 
minimum Lorentz factor occurring in the ejected mass. We then have 
$dF/dx=(\Gamma_0/M_0)dM/d\Gamma$. Note that in general $dF/dx\leq 0$,
since in realistic scenarios we must have 
$\Delta M\geq 0$. Note also that $F(1)=1$. The original fireball scenario
with an instantaneous energy release is equivalent to $dF/dx=0$.

The energy of a mass element $dM$ is $dE=\Gamma dM c^2$, where $c$ is
the velocity of light, and the total
energy of the ejecta with Lorentz factor greater than $\Gamma$ is 
therefore given by 
\begin{equation}
E(\geq \Gamma)= E_0+\Delta E=E_0+\int_{\Gamma_0}^\Gamma
\left( \frac{dE}{d\Gamma}\right) d\Gamma 
=E_0\left(1 +\int_{1}^x x\frac{dF}{dx}dx\right) ,
\label{eq:egtgamma}
\end{equation}
where $E_0=\Gamma_0M_0c^2$. Integrating by parts we find that
\begin{equation}
E=E_0\left( xF(x)+\int_x ^1 F(x)dx\right) .
\label{eq:eintegral}
\end{equation}
The expanding shell decelerates by sweeping up the ambient medium. 
Assuming that the instantaneous total mass of the swept up medium is $M_e$, 
the total energy of the shell
can be expressed as $E=\Gamma(\gamma_i[M(\geq \Gamma)+M_e]c^2)$, where
$\gamma_i$ is the average internal Lorentz factor of the particles in the 
expanding shell.\footnote{To be fully consistent, one would also need to 
distinguish between the internal Lorentz factors of the ejected shells and that of the 
shock-front. This distinction does not, however, 
change the final results as the total energy of the shell cancels out in the 
derivation of equation~(\ref{eq:beta}) below.}
Using this and equation~(\ref{eq:eintegral}) above, we find 
that the conservation of energy may be written as
\begin{equation}
\Gamma\gamma_i(F(x)+f)=\Gamma_0[xF(x)+I_1]+f,
\label{eq:econs}
\end{equation}
with $f=M_e/M_0$ and 
\begin{equation}
        I_1=\int_x^1 F(x)dx. 
\end{equation}
Note that the swept up mass ratio, $f$, increases as $\Gamma$ decreases.  

In the preceding equations we have assumed that the slower moving mass shells
ejected by the central engine catch up with the decelerating shock front as soon 
as the Lorentz factors of the two coincide.  This is a simplification, as the 
speed difference between the ejected mass elements will cause a time
delay.  The dynamical evolution is however not affected by this, but it should 
be kept in mind when considering reverse shocks originating from shells colliding
with the shock front. A method for including this time delay will be discussed 
in section \ref{subsec:discrete}. 

Before impact with the expanding shock front, each mass element
has momentum $dP=\Gamma dM\beta c$, where $\beta c$ is the velocity of the 
mass element in the burster frame.  Hence the total momentum before impact is
\begin{equation}
P=P_0+\int_{\Gamma_0} ^\Gamma dP=P_0-M_0 c\int_x ^1 (\Gamma_0 ^2x^2-1)^{1/2}
\frac{dF}{dx}dx,
\label{eq:totmom}
\end{equation}
with $P_0=\Gamma_0M_0\beta_0c$. As the momentum of the hot shell is 
$P=\Gamma \gamma_i [M(\geq\Gamma)+M_e]\beta c$, momentum conservation
results in 
\begin{equation}
\Gamma\gamma_i\beta (F(x)+f)=\Gamma_0\beta_0-I_2,
\label{eq:momcons}
\end{equation}
where 
\begin{equation}
        I_2 = \int_x^1 (\Gamma_0 ^2x^2-1)^{1/2}\frac{dF}{dx}dx.
\end{equation}
Dividing
equation (\ref{eq:momcons}) by equation (\ref{eq:econs}), we obtain an expression 
for the dimensionless velocity of the expanding fireball,
\begin{equation}
\beta=\frac{\Gamma_0\beta_0-I_2}{\Gamma_0[xF(x)+I_1]+f}.
\label{eq:beta}
\end{equation}
We then find that the Lorentz factor is given by
\begin{equation}
\Gamma=\frac{1}{\sqrt{1-\beta^2}}=
\frac{\Gamma_0[xF(x)+I_1]+f}{\sqrt{[\Gamma_0[xF(x)+I_1]+f]^2-[\Gamma_0\beta_0-I_2]^2}}.
\label{eq:gamma}
\end{equation}
This is an implicit equation for $\Gamma$, and describes the evolution of the
Lorentz factor with radius (or time), once the distribution of the injected
Lorentz factors, $F(x)$, has been specified along with the initial conditions
of the fireball (e.g.\ $\Gamma_0$ and $M_0$). A description of
the swept up mass, $f$, is also required (see eq.~[\ref{eq:dfdr}] below). 
Equation~(\ref{eq:gamma}) is the
key equation for obtaining the dynamical evolution of the fireball and
has to be solved numerically in all realistic cases. Since we take the
evolution to be adiabatic, the dynamics can be solved for separately and the
radiative properties computed afterwards.

For a complete description of the fireball evolution we must solve
four differential equations (R99)
\begin{equation}
        \frac{dr}{dt_b} = \beta c,\qquad
        \frac{dt'}{dt_b} = \frac{1}{\Gamma},\qquad 
        \frac{dt}{dt_b} = \frac{1+z}{\Gamma^2(1+\beta)},\qquad
        \frac{df}{dr} = \frac{\Omega \rho}{M_0}r^2. \label{eq:dfdr}
\end{equation}
Here, $r$ is the radial coordinate in the burster frame and $t_b$, $t'$ 
and $t$ are the burster frame time, the comoving frame time and the observer
frame time, respectively. The solid angle into which the energy
is beamed is estimated by $\Omega = 2\pi(1 - \cos(\theta_0 + c_s t'/(c
t_b)))$ where $\theta_0$ is the initial opening angle of the collimated
outflow and $c_s \approx c/\sqrt{3}$ is the sound speed in the comoving
frame, assumed to be relativistic. The mass density of the external 
medium is $\rho$, which in general is a function of radius. 

\subsection{Instantaneous Energy Injection}
\label{subsec:single}

In the standard fireball scenario it is assumed  that the energy 
is released instantaneously with $x=1$ and $F(x)=1$. We then find that 
$xF(x)+I_1=1$; $I_2=0$ and equation (\ref{eq:gamma}) reduces to
\begin{equation}
\Gamma=\frac{\Gamma_0+f}{\sqrt{[\Gamma_0+f]^2-[\Gamma_0^2-1]}}=
\frac{\Gamma_0+f}{\sqrt{1+2\Gamma_0f+f^2}},
\label{eq:e_inst}
\end{equation}
which agrees with the result of \citet{PaczRhoads93} and is the starting 
point of the R99 analysis (see his eq.\ [4]). Some authors have
based their analysis of GRB afterglows on this equation 
\citep[e.g.][]{Salmonson2003, BiancoRuffini2005}.

GRB~990510 is an example of a burst where the afterglow may be readily
interpreted with a single instantaneous energy release event. We present 
our fit to the observations in Section~\ref{sec:510}.

\subsection{Discrete Energy Injection}
\label{subsec:discrete}
Assume that in addition to the initial injection $M_0$, the central engine 
expels several shells with masses $M_i$ and Lorentz factors $\Gamma_i$.
Converting to dimensionless variables, $x_i = \Gamma_i/\Gamma_0$ 
and $m_i  = M_i/M_0$ we have that
\begin{equation}
        F(x) = 1 + \sum_{i} m_i H(x_i - x),
        \label{eq:fdiscrete}
\end{equation}
where
\begin{equation}
        H(x_i-x) = 
        \left\{ \begin{array}{ll}
              0 & x_i < x\\
              1 & x_i \ge x
               \end{array}
        \right.
\end{equation}
is the Heaviside step function.
We then obtain, $xF(x)+I_1 = 1 + \sum x_i m_i H(x_i - x)$ and
$I_2 = -\sum m_i\sqrt{\Gamma_0^2x_i^2-1} H(x_i-x)$. Note that by the 
assumptions of our approach discussed in the first two paragraphs of 
Section~\ref{sec:dynamics}, the $x_i$'s form a decreasing sequence. 
Inserting equation~\ref{eq:fdiscrete} into equation~(\ref{eq:gamma}) we 
obtain
\begin{equation}
        \label{eq:gammadiscrete}
        \Gamma = \frac{\Gamma_0\left( 1+\sum x_i m_i H(x_i-x) \right) +
        f}{\sqrt{\left[ \Gamma_0\left( 1+\sum x_i m_i H(x_i-x) \right) + f \right]^2
        - \left[ \Gamma_0\beta_0 + \sum m_i \sqrt{\Gamma_0^2x_i^2-1} H(x_i-x)
        \right]^2}}.
\end{equation}
Because the Heaviside function is zero until $x_i\ge x$, it is clear from this 
equation that the dynamical evolution between injections is similar to the instantaneous 
case as the evolution of $\Gamma$ depends only on $f$ in the same way. 

When a shell collides with the shock front, a reverse shock may propagate back
through the shell if the collision is violent \citep{ZhangMesz2002}. Depending 
on the circumstances, radiation from the reverse shock may make a substantial 
contribution to the afterglow light \citep{NakarPiran2004}.
According to \citet{ZhangMesz2002}, the liklihood of getting a reverse 
shock depends on the relative energy and speed between the shells and the width 
of the incoming shell. As the energy is known and we neglect the dynamics in the shell
collision, only the relative speed is needed.
Assuming that only the shock front slows down due to interaction 
with the surrounding medium and that the trailing shell travels with 
constant speed, shells ejected with the average Lorentz factor of the
shock, $\Gamma_a$, will have reached the shock front at $r$ in time $t$.
Replacing $x$ with $x_a=\Gamma_a/\Gamma_0$ in equation~(\ref{eq:gamma}) is then
sufficient to account for the delay.  
As shown by \cite{KumarPiran2000}, the Lorentz factor of the incoming shell 
is $\Gamma_a \approx 2\Gamma$ in a constant density environment for most of 
the evolution.  Combined with the results from 
\citet{ZhangMesz2002} (see their figure 3) this indicates that for a reverse 
shock to result, the energy of the incoming shell must be 
at least twice the energy of the shock front. 

We emphasize that in our numerical calculations we 
assume the shell collisions to be instantaneous and we neglect the 
effects of the interaction on the light curves. These effects are expected to be small
considering that the relative Lorentz factor of the collision is $\Gamma_r
\approx 1.25$ \citep{KumarPiran2000}. We estimate that the shells
would have to be at least several hundred light seconds thick for the
effects of the interaction to be measurable with current instruments;
these would also be strongly smoothed because of the equal
arrival time surface (EATS).

An example of an afterglow that may be interpreted with discrete energy injections 
is that of GRB~021004. \citet{Bjornsson2004} showed that
its properties, including polarization, could be explained
by using such a model. Energy injection does affect the polarization levels 
calculated using the model of \cite{GL1999} as these are most strongly 
determined by the Lorentz factor. Changes in the 
Lorentz factor therefore manifests itself not only in the light curve but 
also in polarization data. An analysis of a more detailed data sets for 
GRB~021004 has been presented by \citet{Postigo2005}. The late time
afterglow and the host galaxy of this burst have been studied by \cite{Fynbo2005}. 

The afterglow of GRB~030329 also showed variations both in the optical
and the polarization light curves. These have previously been interpreted
with energy injections \citep{GraNaPir2003}, although other alternatives have also been
explored \citep{Berger2003, Huang2005}. An application of our approach to that particular 
burst will be presented elsewhere (Guizy et al.~in preparation).

\subsection{Continuous Injection}
\label{subsec:continuous}

Here, we follow the approach of RM98 and SM00 and assume that $F(x)=x^{-s}$,
with $s>1$, i.e.\ the energy is dominated by the lowest Lorentz factors. 
(The case $s<1$ is equivalent to instantaneous injection as the
energy is then dominated by the highest Lorentz factors). 
Then $xF(x)+I_1=Z/\Gamma_0+1$, where 
\begin{equation}
Z=\frac{\Gamma_0 s}{s-1}(x^{-s+1}-1).
\end{equation}
To evaluate $I_2$, we assume that $\sqrt{\Gamma_0 ^2 x^2-1}\approx\Gamma_0x$,
which is always a good approximation if $\Gamma_m ^2\gg 1$.
We then find $I_2\approx -Z$. The dynamical equation for
$\Gamma$~(eq.~[\ref{eq:gamma}]) in this case
becomes
\begin{equation}
\Gamma=
\frac{\Gamma_0+f+Z}{\sqrt{1+2\Gamma_0Z(1-\beta_0)+2\Gamma_0 f(1+Z/\Gamma_0)+f^2}}.
\label{eq:gamma_con}
\end{equation}
Note that for $Z=0$, we recover the result for the instantaneous case (eq.~[\ref{eq:e_inst}]).
Also note that the second term under the square root is $\approx 0$ when $\Gamma_0\gg 1$.
When considering models with continuous injection, we numerically solve 
equation~(\ref{eq:gamma_con}), since in all relevant cases $\Gamma_m^2 \gg 1$.

\section{Simple Approximations} 
\label{sec:approx}

The quantities that determine the dynamical evolution of the fireball
are the dimensionless mass ratios $F(x)$ and $f$, representing the injected 
energy and the swept up mass, respectively (see eq.~[\ref{eq:gamma}]). 
$F(x)$ is an input quantity in the model, and hence each case has to 
be solved separately. On the other hand, $f$ is only dependent on the 
evolution of $r$, $\rho$ and $\Omega$.  In general,
$\rho \propto r^{-g}$, where $g=0$ for constant density interstellar
medium (ISM) and $g=2$ for a wind like environment.  
The evolution of $f$ can easily be obtained in two cases: 
the {\em confined outflow} case when $c_st' \ll \theta_0r$ and $\Omega$ 
may be considered to be a constant. In this case the dynamics is that of
a spherical outflow. The second case is that of a {\em laterally spreading outflow},
when $c_st'\gg \theta_0r$ and $\Omega$ is large enough to neglect the 
initial opening angle. These cases were referred to as the power law regime 
and the exponential regime, respectively, in R99. That naming convention 
is however too limiting as we show below, since for a continuous energy injection 
there is no exponential regime.

As a prescription for $f$ in the confined case, we evaluate it as 
\begin{equation}
f\approx\frac{\Omega_0\rho}{(3-g)M_0}r^3\propto r^{3-g},
\label{eq:ffix}
\end{equation}
while in the laterally spreading case we adopt (see R99)
\begin{equation}
\frac{df}{dr}\approx\frac{\pi}{M_0}\rho(c_st')^2.
\label{eq:dfdrexp}
\end{equation}
To obtain a full solution of the problem and the light curve behavior, 
these need to be consistently included in the calculation.

Approximate solutions applicable in the instantaneous case have been 
discussed in detail by e.g.\ R99 and \citet{Salmonson2003}. We emphasize that the
approximations can give useful insight into the properties of the fireball but they 
should, however, only be expected to be valid during limited time intervals in 
the afterglow evolution. This is partly due to the validity of the underlying
assumptions in those limited intervals, and partly due to the influence of the 
EATS on the evolution (see Section~\ref{sec:EATS}). Another factor is the 
spectral energy distribution (SED), that is not really a set of broken 
power laws (see Appendix~\ref{sec:app}), 
but has an intrinsic curvature, and so do the lightcurves. 
In the following we first give a brief discussion of the discrete case
but then concentrate on the case of continuous energy injection.

\subsection{Discrete injections}

If the energy is released in several discrete events (as discussed in
Section~\ref{subsec:discrete}), the evolution after a shell impact
can be approximated with the evolution following a single release event 
with the appropriate initial parameters. Of these, only the energy is
relevant, being equal to the total energy that has accumulated in the
shock front after the impact. Integration over the EATS then smooths 
out the transition.

As mentioned in Section~\ref{subsec:discrete}, the relative speed
between the impacting shell and the decelerating shock front is an
important factor in evaluating the reverse shock emission arising from
the collision. The speed of the incoming shell equals the average
speed of the shock front at the time of collision, and to estimate it,
we need to know how the radius of the fireball, $r$,
evolves with time in the burster frame, $t_b$.  Using
$\Gamma\approx\sqrt{\Gamma_0/2f}$, appropriate for a single release
(see R99), and equations~(\ref{eq:ffix}) and~(\ref{eq:dfdr}), it is
easy to show that in the confined case the speed of the incoming
shell is $\Gamma_a\approx\sqrt{(4-g)}\Gamma$.  Although this should
only be valid early on in the evolution, numerical calculations
indicate that it is in fact appropriate during most of the evolution.

\subsection{Continuous energy injection}
\label{subsec:coninj}

All results derived in this section assume that the shock speed, $\beta\approx1$. 
We first note that as $f$ is the swept up mass in units of the initial 
ejected mass, it is small compared to the other terms in equation~(\ref{eq:gamma_con}) 
for a substantial fraction of the initial fireball evolution, but will dominate 
at later times. 

Early in the evolution of the fireball, when $x$ is close to 1,
$Z < \Gamma_0$ and we recover equation~(4) in R99, and an identical initial 
fireball evolution. 
The most interesting case, however, is when $Z/\Gamma_0>1$. From the definition
of $Z$ we see that this occurs when $x=\Gamma/\Gamma_0<
\left(s/(2s-1)\right)^{1/(s-1)}$, i.e.\ when $x<2/3$ for $s=2$ and for
$x<\sqrt{3/5}$, when $s=3$.  In this case equation~(\ref{eq:gamma})
simplifies to
\begin{equation}
\Gamma\approx\frac{\Gamma_0+f+Z}{\sqrt{1+2Zf+f^2}}.
\end{equation}
When both $\Gamma_0$ and $f$ can be neglected compared to $Z$, we have that
$\Gamma\approx \sqrt{Z/2f}$. Dividing by $\Gamma_0$ and solving for
$x=\Gamma/\Gamma_0$,
we find 
\begin{equation}
x\approx\left(\frac{1}{2\Gamma_0}\frac{s}{s-1}\right)^{1/(s+1)}f^{-1/(s+1)}.
\label{eq:x}
\end{equation}
Note that in the limit, $s\rightarrow 1$, we recover the power law behavior 
of the instantaneous injection case, where $x\propto f^{-1/2}$ (R99).

{\em Confined outflow:} Here, we adopt equation~(\ref{eq:ffix}), with 
$\Omega_0\approx\pi\theta_0^2$, as the solid angle into which the beam 
propagates. Inserting into equation~(\ref{eq:x}), we find that
\begin{equation}
x\approx\left(\frac{1}{2\Gamma_0}\frac{s}{s-1}\frac{(3-g)M_0}{\Omega_0\rho}\right)^{1/(s+1)}r^{-3/(s+1)}
\propto r^{-(3-g)/(s+1)}.
\label{eq:xx}
\end{equation}
As in R99, we also evaluate the relevant times. From $dt'/dr=1/(c\Gamma_0x)$, 
we find by integration that the time in the comoving frame is
\begin{equation}
t'\propto \left(\frac{s+1}{s+4-g}\right)r^{(s+4-g)/(s+1)}.
\end{equation}
Similarly, from $dt/dt_b=(1+z)/(2c\Gamma_0^2x^2)$,
the time in the observer frame is given by 
\begin{equation}
t\propto\left(\frac{s+1}{s+7-2g}\right)r^{(s+7-2g)/(s+1)}.
\end{equation}
Inverting this result and inserting into equation~(\ref{eq:xx}), we find that
$\Gamma\propto t^{-(3-g)/(s+7-2g)}$, in agreement with the SM00 result.
Also note that if we set $s=1$ and $g=0$, we recover the 
results of R99.

{\em Laterally spreading outflow:} The treatment is similar to that of R99. The 
$g=2$ case cannot, however, be solved analytically due to the coupling of the 
variables. We therefore only consider the case $g=0$ in this section (constant 
density ISM). The general case has to be solved numerically. 

We start from equation~(\ref{eq:dfdrexp}), divide by $dt'/dr$ and integrate to obtain
\begin{equation}
f^{(2+s)/(1+s)}\propto
\frac{\rho c_s^2}{3}({t'}^{3}-c_1),
\label{eq:fs}
\end{equation}
in agreement with equation~(14) of R99 if we set $s=1$. Here, $c_1$ is a constant 
of integration to be determined by the initial conditions. It becomes
negligible when $c_st' \gg \theta_0 r$.

The exponential behavior discussed by R99 is seen to follow from
equation~(\ref{eq:fs}).
If $s=1$ (formally equivalent to the instantaneous case), then equation~(\ref{eq:fs})
shows that $f\propto {t'}^{2}$ so that $dt'/dr\propto t'$, and
the time in the comoving frame increases exponentially with $r$.
Therefore $f$ also increases exponentially with $r$.

On the other hand, if $s>1$, the behavior becomes a pure power-law,
$f\propto {t'}^{3(1+s)/(2+s)}$ and $t'\propto r^{(2+s)/(s-1)}$
which shows that the exponential regime is a special case. 
It follows that $f\propto r^{3(1+s)/(s-1)}$,
$r\propto t^{(s-1)/(5+s)}$ and $\Gamma\propto t^{-3/(5+s)}$.
The physical reason for this change in behavior is that the 
dynamical evolution is now determined by the power-law energy 
injection rather than the lateral spreading. We note that in
general, if the incoming shell has the same opening angle as 
the initial $\theta_0$, the resulting decay will be a mixture
of a power-law decay for the part of the outflow within $\theta_0$, 
and an exponential for that part of the flow that is outside 
of $\theta_0$.

By assumption, the dynamical evolution is independent of the radiative 
properties of the fireball. To proceed, a prescription of the radiative 
processes is required. We adopt the standard synchrotron emission 
mechanism \citep[see e.g.][for a review]{Piran2005}, that is widely assumed
to be the source of radiation in GRBs \citep[e.g.][]{Tavani1996, MeszRees1997,
SarPirNar1998}. The details of the radiative component of our modeling are 
given in Section~\ref{sec:lcurves} and in Appendix~\ref{sec:app}.

To explore the approximate light curve behavior, $F_\nu \propto t^{-\alpha}\nu^{-\beta}$,
we need to evaluate the characteristic frequencies $\nu_m$, $\nu_c$ and the
flux $F_{\nu_m}$ at the former (See Appendix~\ref{sec:app} for definitions of these
quantities). 

For the confined case, the temporal behavior of the characteristic frequencies, 
the $F_{\nu_m}$, and the light curve decay indices, $\alpha$,
agree with those previously derived for the forward shock by SM00 
(see Table 1 in their paper). 

For the laterally spreading case we find on the other hand
\begin{eqnarray}
\Gamma & \propto & t^{-3/(s+5)}, \\
r      & \propto & t^{(s-1)/(s+5)},\\
\nu_m  & \propto & t^{-12/(s+5)},\\
\nu_c  & \propto & t^{-2(s-1)/(s+5)},\\
F_m    & \propto & t^{(3s-9)/(s+5)}. 
\end{eqnarray}

\begin{table}[t]
        \centering
        \caption{Exponents for $F_\nu \propto t^{-\alpha}\nu^{-\beta}$ for a 
          laterally expanding beam.
}
\vspace{5mm}
                \begin{tabular}{|c|c|c|}
                \hline
                & $\nu_m<\nu<\nu_c$ & $\nu>\nu_c$\\
                \hline                                             
                $\alpha$ & $3(3-s+4\beta)/(s+5)$ & $2(1-s+6\beta)/(s+5)$ \\
                $\beta$ & $ (p-1)/2           $ & $p/2$ \\
                \hline
                \end{tabular}
        \label{table:exps}
\end{table}
\noindent
The corresponding light curve exponents are given in
Table~\ref{table:exps}. (Note that if we let $s=1$ we recover the
corresponding results for the instantaneous case). From the 
indices in Table~\ref{table:exps} and the results of SM00 we see
that if $\nu_m<\nu<\nu_c$, the steepening in the light curve when the
observer starts to see the entire outflow surface is
$\Delta\alpha=\alpha_2-\alpha_1 =24(2+\beta)/((s+5)(s+7))$, with
$\beta=(p-1)/2$. For parameters typically inferred in afterglows ($p=2.5$
and $\beta \sim 0.75$) and taking $s=2$ as an example, we see that $\Delta\alpha\sim 1$. 
When $\nu>\nu_c$ the result is $\Delta\alpha=24(1+\beta)/((s+5)(s+7))$,
with $\beta=p/2$ in this case. For typical parameters we now find $\Delta\alpha\sim 0.85$. 
As the steepening of the light curve is similar in the two cases,
the magnitude of the break does not indicate in which spectral region the
observing frequency is. Clearly, a more detailed fit to the observations 
is needed. Inferring $p<2$ from a standard model fit may be taken as a strong
indication that a continuous injection may be involved.

\section{Light Curves and Spectra}
\label{sec:lcurves}

As is customary in GRB fireball models and discussed in the previous
section, we assume the radiation to be of synchrotron origin.  We
adopt the approach of \citet{Wuetal2004}, although their method of
calculating the radiation flux by integrating the synchrotron
emissivity over the electron energy distribution is however too
computer intensive for our purposes. Instead, we use a broken power
law approximation as in \citet{WijersGalama1999}, smoothly joined
together similar to \citet{GranSar2002}.  The detailed expressions are
summarized in Appendix~\ref{sec:app}.

Similarly to \citet{Wuetal2004}, we calculate the synchrotron self absorption 
coefficient directly and therefore the self absorption frequency is not included 
in our smoothly joined power law approximation. We also include inverse Compton 
scattering, but neglect higher orders as these are suppressed by the Klein-Nishina 
effect \citep{PanaitescuKumar2000}. The Compton $Y$ parameter is calculated from 
\begin{equation}
        Y =
        \frac{4}{3}\tau_e\int_{{\cal N}_e}\gamma^2{\cal N}_e(\gamma)\,d\gamma,
\end{equation}
where ${\cal N}_e$ is the normalized electron distribution and
$\tau_e$ is the optical depth to electron scattering.  The modified cooling
Lorentz factor is found from the implicit equation
\begin{equation}
        \gamma_{c, IC} = \frac{\gamma_c}{1+Y},
\end{equation}
where, $\gamma_c$, is the usual cooling Lorentz factor (see eq.~\ref{eq:gamma_c}).  
The Comptonized emission component is then found from the synchrotron spectrum 
using the scalings,
$\nu_m^{IC} = 2\gamma_m^2\nu_m$, $\nu_c^{IC} = 2\gamma_{c, IC}^2\nu_{c, IC}$ and 
$P_{{\rm max}}^{IC} = \tau_e P_{{\rm max}}$  \citep{SariEsin2001}.

\subsection{The emitting region}
\label{sec:EATS}

In the standard afterglow model the fireball is described by the
self-similar formulation of \citet{BM1976} (hereafter BM). Analytical 
solutions are however only available if the ambient medium has a density 
profile of the form $n\propto r^{-g}$. Some authors, 
\citep{NakarPiran2003,Lazzati2002}, have adopted these solutions in models 
with modest density variations. Using their approach, we are able to 
reproduce their results only if we assume that the shock thickness 
follows the BM solution, $\Delta \sim r/\Gamma^2$. 
Care must be exercised 
here, since in this approximation the number of radiating particles may not 
be estimated correctly. To clarify, let us consider a fireball in a constant 
density ISM. Using the \citet{BM1976} solution for the density behind the 
shock front and integrating over the swept up volume we find that the number 
of particles contained behind the shock is 
\begin{equation}
        N_{{\rm p}} = \frac{4\pi r^3n}{3} + O(\Gamma^{-2}),
\end{equation}
where $r$ is the the radius of the swept up volume, 
and $n$ is the (constant) ISM particle density.  This result is appropriate for a 
homogeneous medium but leads to over- or underestimates, depending 
on the sign of the density gradient, in the number of radiating 
particles when used for variable ISM densities. In such cases it 
leads to incorrect flux estimates.  We also remind the reader that 
density variations do affect the evolution of $\Gamma$ which has to 
be calculated self-consistently to obtain correct fluxes.

Instead of using the BM solution directly, we make the simplifying assumption
that the shock front is homogeneous in the comoving frame. Using particle 
conservation and the jump conditions we find that the shock thickness in
the burster frame is 
\begin{equation}
        \Delta = \frac{M_e}{2\pi m_p r^2(1-\cos\theta)\Gamma n'},
\end{equation}
appropriate for all density profiles. Here, $m_p$, is the proton mass and,
$n'=4\Gamma n$, is the density in the comoving frame. Note that when
$(1-\cos\theta)=2$, and $M_e=4\pi r^3n/3$, we recover the BM shock thickness.
Although apparently a crude estimate, \citet{GranPirSar1999} showed 
that the difference between the calculated
flux levels in such a thin shell approximation and
the full solution are small.

In all our flux calculations, we follow the evolution of the
EATS. The total flux arriving at the 
observer at a given time, $t$, is obtained by integrating over the EATS for
a given observer frequency, $\nu$, \citep{GranPirSar1999}:
\begin{equation}
        F(\nu,t) = \frac{1+z}{2d_l^2}\int_0^\theta \int_{0}^r 
                                \frac{P'(\nu',t_{{\rm em}},\vec{r})}{\Gamma^2(1-\beta\cos\theta)^2}
                                r^2 dr\; d\cos\theta,
\end{equation}
where, $P'$, is the radiation power density in the comoving frame (see also appendix~\ref{sec:app}),
$\nu'$ is the frequency in the comoving frame and
\begin{equation}
        t_{{\rm em}} = \frac{t}{1+z} + \frac{r\cos\theta}{c},
\end{equation}
is the time of emission in the burster frame. Also, $d_l$,
is the luminosity distance which depends on the cosmology. In our numerical
calculations we assume a standard cosmological model
with $\Omega_\Lambda = 0.7$, $\Omega_m = 0.3$ and $H_0 = 70$
km s$^{-1}$ Mpc$^{-1}$.  

\subsection{Energy injection effects on radiation properties}
\label{sec:radprop}

To understand the effects energy injections have on the radiation
properties of the afterglow it is best to start by considering the
effects of a single injection. Since we neglect the
dynamics of shell collisions, the Lorentz factor of the leading shell
increases instantaneously as soon as another shell catches up with it.  The
subsequent evolution will then quickly relax to a solution which is similar 
to the one obtained if the total energy of both shells was injected instantaneously.
As the energy of the shock front increases instantaneously, so does its luminosity.
However, the increase in flux received by the
observer will not be instantaneous but gradual because the integration over
the EATS smooths out the transition. The timescale of this gradual
brightening depends highly on both the Lorentz factor and the jet
opening angle \citep{Piran2005}.  Energy injection in wider and slower moving jets 
will cause the flux to change on longer timescales and thus creates 
smoother bumps in the light curves. Shell interactions considered 
in more detail than in this work, would also contribute to the smoothening.

\section{Examples of data modeling}
\label{sec:interp}

In this Section we apply our model to observations and give three examples 
of how well our numerical results fit afterglow data. We use 
reduced $\chi^2$ deviation to assess the goodness
of the fit. As our model extends to the entire electromagnetic spectrum, 
an un-weighted $\chi^2$ is not adequate for our purposes.
The number of points in the optical light curves often exceeds that in other 
wave bands by a factor of ten or more which tends to decrease the quality 
of the fit at the radio and X-ray wavelengths. To correct
for this, we weight the contribution from each band with the square root of the
number of data points in that band. We then normalize the result so that if 
the contribution from each point to the un-weighted $\chi^2$ is unity, the 
weighted $\chi^2$ equals the total number of data points.
The function we use to check the goodness of the fit is then
\begin{equation}
        \label{eq:goodness}
        \chi^2_{{\rm d.o.f}} = \frac{1}{\left(N_{{\rm dp}}-N_{{\rm par}}\right)}
        \frac{N_{{\rm dp}}}{\sum_{i=1}^{N_{{\rm lc}}}\sqrt{N_i}} 
        \sum_{i=1}^{N_{{\rm lc}}}\frac{1}{\sqrt{N_i}}
        \sum_{j=1}^{N_i}\left(\frac{\log(F_{i,j})-\log(F_i(t_j))}
        {\log\left(1 + \Delta F_{i,j}/F_{i,j}\right)}\right)^2,
\end{equation}
where $N_{{\rm dp}}$ is the total number of data points, $N_{{\rm lc}}$ 
is the number of light curves, $N_i$ is the number of data
points for light curve $i$, $N_{{\rm par}}$ is the number of
parameters, $F_{i,j}$ is the flux in light curve $i$, measured at time
$t_j$ with the error $\Delta F_{i,j}$, and $F_i(t_j)$ is the flux from
our model light curve $i$ at time $t_j$.  Dividing by the number of
degrees of freedom (d.o.f.) in the first factor of the equation provides 
the reduced $\chi^2$ which we denote by $\chi^2_{{\rm d.o.f.}}$. 
The second factor is the normalization, and $1/\sqrt{N_i}$ in the sum 
over light-curves (lc) is the weight.  Note that the
fit is done for $\log(F)$ rather than $F$, as this is equivalent to a
liner fit in magnitudes.

Fitting the model calculations to the measurements is nontrivial
since the number of model parameters can be large and the goodness of fit
function, $\chi^2_{{\rm d.o.f}}$, has multiple local minima. This makes local gradient
based search methods unsuitable for our purposes. Instead we employ a
global search method known as the Evolution Strategy \citep{Thomas2000}.
The essence of the algorithm is as follows: Initially, a
population of randomly generated $\lambda$ search points (parameter
sets) are generated within the specified parameter bounds and the
initial variance of the normally distributed search distribution is
set. These $\lambda$ points are then evaluated using the $\chi^2_{{\rm d.o.f}}$
function and sorted. The best $\mu$ of those are then used as parent search
points and each one of them generates $\lambda/\mu$ new points within
their Gaussian search distribution. However, before doing so the
search distribution itself, or rather its standard deviation, is
varied using the log-normal distribution. In this manner the most
favorable search distribution is adapted to the function landscape
resulting in faster search. Initially the search distribution reaches
the entire search space.  The evolutionary process is repeated for a
number of iterations until the search distribution's standard
deviation is below a pre-specified limit.

In our fits, we did not allow the $g$-parameter (the ambient density 
profile) to vary, but assumed it to be fixed at either $g=0$ 
(homogeneous ISM) or at $g=2$ (wind). In the following sections, 
we set $g=0$, as this provides a better fit in 
all three cases considered here. In addition, the angle that the line 
of sight makes with the jet axis (the viewing angle) is assumed 
to be zero. This is because it mainly affects the polarization
properties of the afterglow that are not considered here.

The complex landscape of the goodness function, caused by the non
linearity and complexity of the underlying model, makes it difficult
to find a unique parameter set that is a well defined global minimum
of the goodness function. Degeneracy of the parameters can cause
several minima to be considered as candidates for the global minimum.
The $1-\sigma$ errors we quote in our parameter determinations reflect
the statistical uncertainty of the model parameters as determined from 
the measurements and do not include estimates accounting for
uncertainties in the model 
\citep[see also][for a similar discussion]{Panaitescu2001}.  
The results of our fits for three GRB afterglows
are given in Table~\ref{tab:fitpar}.  Several comments on the results
are in order.  As the afterglow light curves are weakly dependent on
$\Gamma_0$, it is not constrained by the data. We only quote
lower limits on it.  The electron energy slope, $p$, is constrained
both by the spectral slopes and the afterglow slopes and is the most
constrained model parameter.  In the case of GRB~000301C we only quote an
upper limit on $\epsilon_B$, as it is poorly constrained, mostly due
to the lack of $X$-ray data.  Finally, we note that the $1-\sigma$
errors quoted in the table are generally not symmetric around the best
fit values.

\subsection{GRB~990510}
\label{sec:510}
GRB~990510 has well resolved afterglow light curves in many wavelengths
and shows a smooth achromatic break during the second day of the
afterglow \citep{Harrison1999,Kuulkers2000,Stanek1999, Holland2000}.  
This afterglow is a good example of smooth lightcurves and can be
interpreted within the standard fireball model
\citep{PanaitescuKumar2001b}.  It is therefore an ideal burst to test
the accuracy of our model and to compare its results to those obtained
by others.

\begin{figure}[t]
\epsscale{0.7}
        \begin{center}
                \plotone{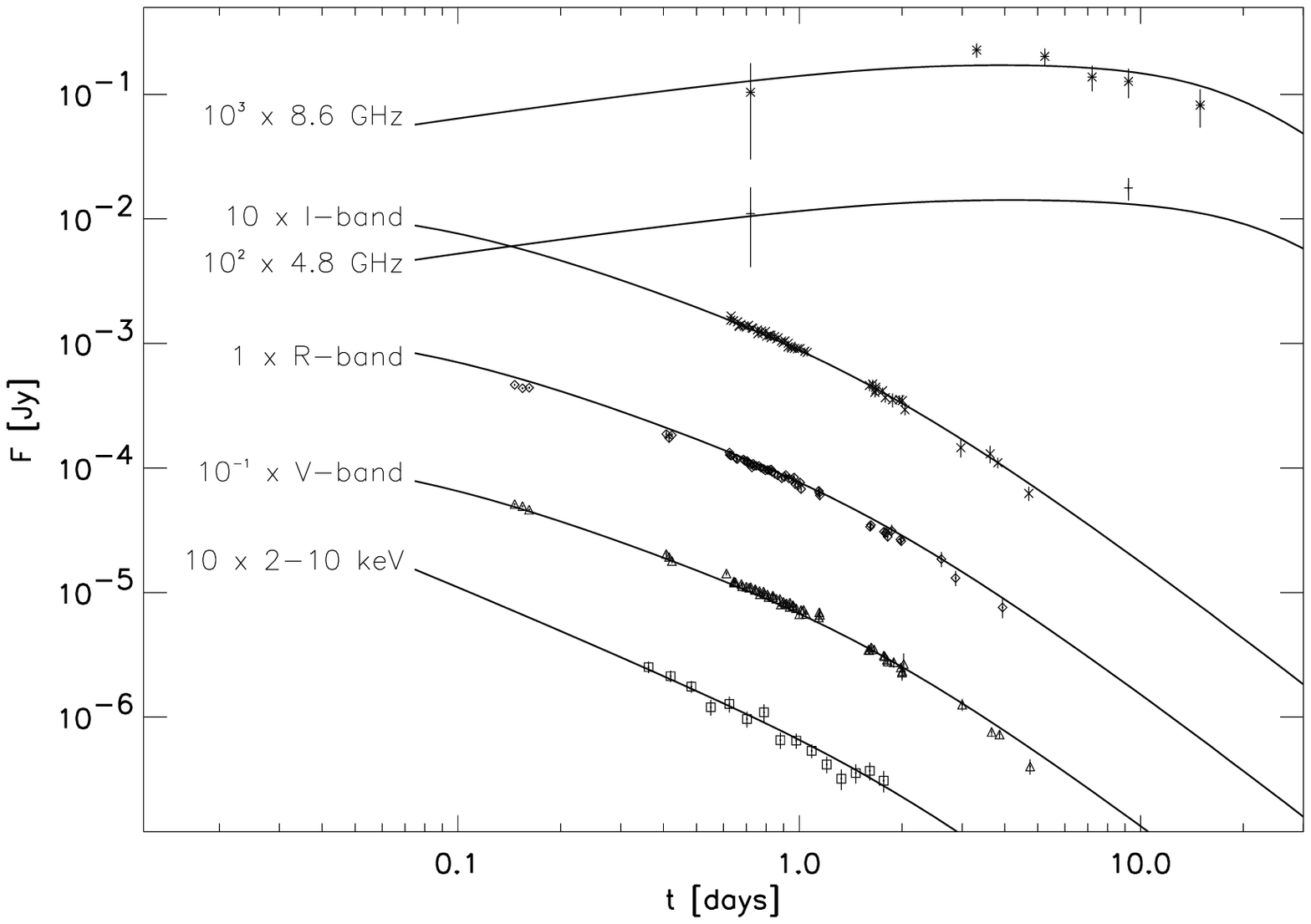}
        \end{center}
\vspace{-0.3cm}
        \caption{The afterglow light curves of GRB~990510 fitted
                     with a model with an instantaneous energy release. The 
                     multi-band fit is reasonably good, resulting in
                     $\chi^2_{{\rm d.o.f.}}=2.6$ with the model parameters given in
                     table~\ref{tab:fitpar}. The X-ray data is from \citet{Kuulkers2000}, 
                     the optical data from \citet{Harrison1999,Stanek1999} and the radio data from
                     \citet{Harrison1999}.  The optical data is corrected for Galactic
                     extinction of $E(B-V)=0.2$ \citep{Schlegel1998}.}
        \label{fig:990510}
\vspace{-0.5cm}
\end{figure}

Figure~\ref{fig:990510} shows the afterglow light curves in radio,
optical and X-rays.  The fit resulted in $\chi^2_{{\rm d.o.f.}} = 2.6$,
quite an acceptable fit given the simplifying assumptions for the
underlying dynamics and radiation and the fact we are fitting the
entire electromagnetic spectrum with a single model.  The best fit parameters
are given in table~\ref{tab:fitpar}. Most of them are similar to the parameters 
found by \citet{PanaitescuKumar2001b}, within a factor of two or three. 
The order of magnitude difference in $\epsilon_e$, the fixed fraction of 
the total internal energy assumed to be in the electron population, stems
from our different definition of $\gamma_m$. Although not shown here, 
the slope of the light curves vary continuously from $\alpha=0.6$ at 0.1 day to 
about $\alpha=2.2$ at 30 days. A rapid steepening sets in at 1 day after the burst
but there is no sharp transition from a pre- to a post-break slope.
We note that for this burst, $\theta\approx 1/\Gamma$ at 0.7 days, just
before the steepening of the light curve.

\begin{table}[t]
        \centering
        \caption{Values of parameters obtained by fitting the three
        afterglows. The errors quoted are the 1-$\sigma$ estimates.}
\vspace{3mm}
        \begin{tabular}{c|c|c|c}
         Parameter                       & 990510  & 000301C & 010222 \\
        \hline                                             
    $\chi^2_{{\rm d.o.f.}}$         & 2.6                             &   3.8                       &  4.6   \\
    $E_{{\rm tot}}$ [10$^{50}$ erg] &  1.8                            & $52$                        & 17   \\
    $E_0$ [10$^{50}$ erg]           & $1.8_{-0.3}^{+0.9}$             & $16\pm4$                    & $1.3_{-0.4}^{+0.2}$   \\
    $\Gamma_0$                      & $>300$                          & $>500$                      & $>800$   \\
    $n_0$ [cm$^{-3}$]               & $4.1_{-1.8}^{+3.3}\cdot10^{-3}$ & $0.18_{-0.15}^{+0.36}$      & $3.4_{-2.7}^{+6.1}\cdot10^{-3}$\\
    $\theta_0$                      & $(1.2\pm 0.3)^{\circ}$          & $1^{\circ}.1_{-0.4}^{+0.5}$ & $1^{\circ}.1_{-0.3}^{+0.2}$ \\
    $p$                             & $2.08_{-0.03}^{+0.14}$          & $2.43\pm0.16$               & $2.22\pm0.05$ \\
    $\epsilon_e$                    & $0.25_{-0.09}^{+0.25}$          & $0.15_{-0.04}^{+0.10}$      & $0.11_{-0.06}^{+0.03}$ \\
    $\epsilon_B/10^{-4}$            & $3.5_{-1.9}^{+2.5}$             & $<4.0$                      & $4.5_{-3.6}^{+1.3}$ \\
    $s$                             &                                 &                             & $1.47_{-0.38}^{+0.04}$ \\
    $\Gamma_m         $             &                                 &                             & $62_{-12}^{+10}$   \\
    $E_1$ [$E_0$]                   &                                 & $2.1\pm0.8$                 &        \\
    $t_1 [{\rm days}]$              &                                 & $1.4_{-0.2}^{+0.1}$         &        \\
        \end{tabular}
        \label{tab:fitpar}
\end{table}

GRB~990510 was the first burst where a polarized afterglow was detected
\citep{Covino1999}. However, as there was only one positive detection
it does not constrain the model very much, although it further strengthens 
the case for a synchrotron origin of the afterglow emission. This 
polarization measurement is easily accounted for by an off-axis viewing
angle. It was not included in the fit presented here.

\subsection{GRB~000301C}
The afterglow of GRB~000301C is well time resolved at optical and radio
wavelengths and was observed in the near infrared and at mm wavelengths
although no $X$-ray observations have been published. A short lived 
brightening was observed in the optical light curves at about 3.5 days
followed by a steep achromatic break around 7 days after the burst
\citep{Sagar2000,Berger2000}.  It has been suggested that this bump 
may be caused by micro-lensing \citep{Garnavich2000}. Their $\chi^2$ value 
of the fit is reduced by half by introducing the lens effect.  

\begin{figure*}[t]
\epsscale{0.7}
        \begin{center}
          \plotone{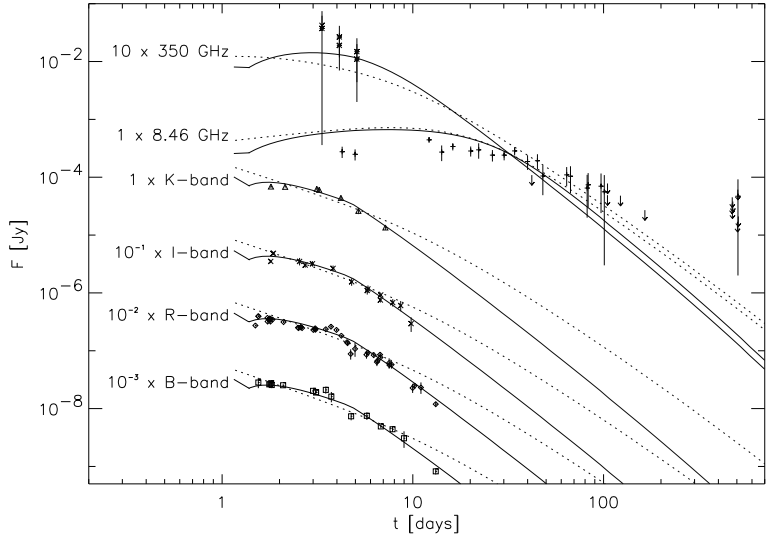}
        \end{center}
        \caption{A sample of the available afterglow light curves of
          GRB~000301C fitted with a jet profile and an additional
          energy injection, $E_1=2.1 E_0$ at $t_1=1.4$ days after the
          burst (solid curves). The fit has $\chi^2_{{\rm d.o.f.}} =
          3.8$ using all available data and is a reasonably good fit.
          The energy injection allows for both the sharp break in the
          optical/nIR bands and the smooth decline of the radio
          light-curve. The bump around 4 days seen in the $R$ band
          causes some of the contribution to $\chi^2_{{\rm d.o.f.}}$
          along with the radio data. It rises and decays too sharply
          to be fitted with an energy injection.  The model parameters
          are given in Table~\ref{tab:fitpar}.  The dotted curves show
          the best fit model without additional energy injection,
          resulting in $\chi^2_{{\rm d.o.f.}} =11$. See text for
          further discussion.  Optical and near infrared data are from
          \citet{Jensen2001,Rhoads2001,Bhargavi2000,Masetti2000} and
          \cite{Sagar2000}. Radio and mm data are from
          \citet{Berger2000,Smith2001} and \cite{Frail2003}. All
          optical data is corrected for Galactic extinction of
          $E(B-V)=0.053$ \citep{Schlegel1998} and intrinsic SMC like
          extinction of $A_B = 0.1$ using the profile of
          \citet{Pei1992}.  A host contribution of $40\pm20$~$\mu$Jy
          has been subtracted from the 8.46~GHz data.}
\label{fig:000301C}
\vspace{-0.5cm}
\end{figure*}

Here, we fit the afterglow light curves with our model containing one additional 
energy injection, $E_1$ at $t_1=1.4$ days post-burst. This is shown in 
figure~\ref{fig:000301C}, where a host contribution of $40\pm 20$ $\mu$Jy at 8.5~GHz has 
been subtracted from the data. The data in the figure is corrected for Galactic 
extinction of $E(B-V) =0.053$ \citep{Schlegel1998}, and an intrinsic SMC like 
extinction of $A_B = 0.1$ using the profile of \citet{Pei1992}.  
This value for the intrinsic extinction was found by \citet{Jensen2001} which
fitted a power law spectrum with various extinction curves to the measured SEDs.

Fitting the data with a model with just an initial instantaneous
energy release results in $\chi^2_{{\rm d.o.f.}} = 11$ (dotted curves
in fig.~\ref{fig:000301C}).  Adding the energy injection $E_1$,
reduces this by a a factor of three to $\chi^2_{{\rm d.o.f.}} = 3.8$,
and gives the parameters presented in table~\ref{tab:fitpar}. Most of
the reduction in the $\chi^2_{{\rm d.o.f}}$ is due to the optical light
curves, as the energy injection allows for a fairly constant emission
at 8.5 GHz and the sharp steepening of the optical light curves at
approximately 6-7 days (solid curves in fig.~\ref{fig:000301C}).  The
remaining contribution to the $\chi^2_{{\rm d.o.f}}$ is due to the
bump at 3.5 days in the optical light curves, that rises too sharply
to be fitted with an injection event. This is because of the EATS and
the homogeneity of the emitting region in our model, making the flux
increase too smooth to fit the sharp rise and decline observed.  This
sharp bump may be due to an injection that is local in the sense that
it only affects a small patch of the radiating shock front. We also 
note that the best fit is obtained by having the energy injection at
$t_1=1.4$, i.e.\ just before the first optical data points, rather
than at times just before the light curve bump. Such an early injection 
does fit the flux levels in the various bands better than an injection
that is constrained to lie within the time window of the observations.

Our model parameter values agree with those of \citet{Panaitescu2001} 
within a factor of 2 or 3, except for quantities that are defined differently
(such as $\epsilon_e$).  Given the differences in approach and the
different level of details in his work and ours, these differences are
rather minor. 

Finally, although energy injection that is uniform across the forward
shock front is unable to account for the bump in the lightcurves at 3.5 days, 
other explanations may apply besides micro-lensing. These could be density 
inhomogeneities in the ambient medium or an inhomogeneous injection
that energizes the emitting surface only partly (patchy shell).

\subsection{GRB~010222}
GRB~010222 had well time resolved afterglow light curves at most wavelengths
and can be interpreted with the standard broken power law in the optical bands.  
The shallow slopes of the light curves require a hard electron energy distribution 
($p<2$), if interpreted within the standard model. An unknown host extinction
may influence the SED slope that in turn may lead to an inconsistent determination 
of $p$, between the spectra and the light-curves.

\begin{figure}[t]
\epsscale{0.7}
        \begin{center}
          \plotone{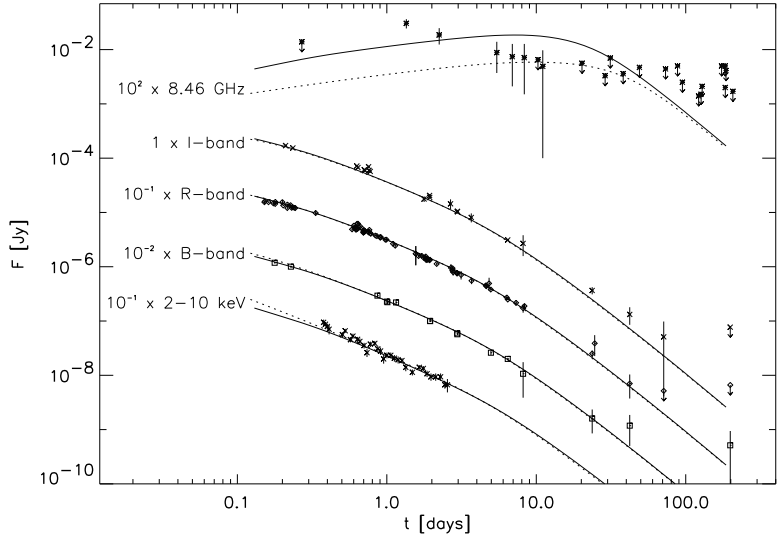}
        \end{center}
        \caption{A sample of the available afterglow light curves
          of GRB 010222 fitted with a jet profile and a
          continuous energy injection (solid curves).  The data has been corrected
          for Galactic extinction.  The fit gave $\chi^2_{{\rm
              d.o.f.}} = 4.6$, using all available data where the
          radio data gives most of the contribution. The model
          parameters are given in table~\ref{tab:fitpar}.  Optical
          data are from \citet{Galama2003,Cowsik2001,Sagar2001} and
          \cite{Stanek2001}. X-ray date from \citet{intZand2001} and
          radio and millimeter data from \citet{Frail2003} and
          references therein. The dotted curves show for comparison a
          fit with only instantaneous injection, resulting in
          $\chi^2_{{\rm d.o.f.}} = 6$.  The increase in $\chi^2_{{\rm
              d.o.f.}}$ is mostly due to contributions from the radio
          light curve.}
\label{fig:010222}
\end{figure}

\citet{Bjornsson2002} found that a continuous energy injection in an
otherwise standard fireball model could explain both the light curves
and SEDs without a hard electron energy distribution.  We follow their
suggestion and use continuous energy injection model to fit the
afterglow data.  The result is shown in figure~\ref{fig:010222} where
the data points have been corrected for Galactic extinction of
$E(B-V)=0.018$ \citep{Schlegel1998} and intrinsic SMC like extinction
of $A_B = 0.1$ using the profile of \citet{Pei1992}, that 
was also adopted by \citet{Galama2003}.  The host of
GRB~010222 is a bright starburst galaxy \citep{Frail2002} and where 
available, the host contribution has been subtracted from the light curves.

The fit resulted in $\chi^2_{{\rm d.o.f.}} = 4.6$ with the model parameters
presented in table~\ref{tab:fitpar}.  The fit is not particularly good, mainly 
because the model gives too high a flux at the radio wavelengths. Both the 
optical and the X-ray data are well accounted for.  We note that the value 
of $\Gamma_m$ is rather high (table~\ref{tab:fitpar}) and the 
continuous energy injection ceases at approximately $0.1$ day, which along with
the jet geometry ($\Gamma\approx 1/\theta$ at about 3 days), accounts for 
the steepening in the light curves.
If we fit the data without the continuous energy injection, we obtain 
$\chi^2_{{\rm d.o.f}}= 5.8$ instead of 4.6, with the additional contribution 
mainly due to a worse fit at radio wavelengths.
The model parameters, however, are similar in both cases; a narrow 
and energetic jet with a low magnetic field strength in a low density 
environment. The continuous energy injection thus better accounts for the
radio light curves but otherwise the evolution in both cases is similar. 
We are therefore able to fit the optical data reasonably well within the 
standard model and still have $p > 2$.

\section{Discussion}
\label{sec:conc}

We have presented a detailed description of a relativistic fireball
afterglow model with discrete and continuous energy injection. We have
given explicit examples of fits to observations for both types of
injections and shown how they can provide a better fit to afterglow
measurements than the standard model with single initial energy
release.  

Although not a direct subject matter of the present paper, we note 
that even polarization measurements can in some cases be accounted for
with the extended model~\citep{Bjornsson2004}. However, interpreting afterglow
polarimetry is not straightforward.  First, polarization measurements
are difficult as the polarization level in general seems to be small
in afterglows.  The data may therefore be sensitive to polarized
contributions from other sources such as the host galaxy, which however
is most likely to be constant. Second, our model is sensitive to the
geometry of the outflow and the assumed homogeneity of the emission
region.  The expanding shock front is unlikely to be
homogeneous during its entire evolution, especially with the added
energy injections.  This is because an injected shell will not
expand laterally in exactly the same way as the original shock front.
The transverse size of the incoming shell will then be smaller and its 
energy deposition in the shock front will be concentrated in the contact 
zone between the two. Accounting for the effects of the EATS, could then 
lead to a rise in the level of polarization shortly after the injection 
and even abrupt changes in the polarization angle if the center of the
colliding shell does not coincide with the geometrical center of the
front.

The origin of the additional shells is most likely the same as the origin
of the shells producing the gamma ray burst in internal shocks.  It is
highly unlikely that the matter ejected from the central engine always
collides and collects into one shell in the gamma ray burst event.  The
remaining shells will then be the ones to refresh the external shock
later on in the evolution of the fireball.  This might even be the
case in most bursts, but goes unnoticed in many cases since the energy
of the colliding shells must be of the same order or higher as the
total accumulated energy of the front if their interaction is to 
produce a measurable brightening.

When fitting GRB afterglow light curves with the standard fireball
model, it is quite common to use power law approximations with a light
curve break due to jet structure.  The slope of the light curve,
$\alpha$, before and after the break is then used to determine the
electron energy distribution index, $p$, which is then compared to the
value of $p$ inferred from the SED of the burst (see e.g.\ Sari et al.
1998).  These results must, however, be carefully considered, as the
value of $\alpha$ depends on $s$ as well (see table~\ref{table:exps}),
and may not be smoothly varying between two fixed asymptotes,
although that seems to be the general case. This is especially true when 
the characteristic frequencies of the synchrotron spectrum, $\nu_m$ and
$\nu_c$, are near the observing frequency, where the SED does not
follow a power law.

The light curve break time has traditionally been used to infer the
jet opening angle from the assumption that it occurs when $\theta\sim
1/\Gamma$. This can lead to erroneous angle estimates, because energy
injections can delay the appearance of the light curve break. An example 
of this is GRB~021004 \citep{Bjornsson2004}. As a result, the 
opening angle will be overestimated and the energy corrected for beaming 
may thus not be correct \citep{Johannesson2006}. Studies based on such energy 
estimates \citep[e.g.][]{Frail2001, Bloom2003, Ghirlanda2004}, should 
include error analysis that accounts for this 'bias'.

\acknowledgements
{We thank P.\ Jakobsson for critical comments on the manuscript and
T.\ P.\ Runarsson for helpful discussion on the fitting procedures. 
We thank the anonymous referee for detailed and constructive criticism.
This work was supported in part by a Special Grant from the Icelandic
Research Council, by the University of Iceland Research Fund and by the
Graduate Research Fund of the Icelandic Centre for Research.}

\appendix
\section{Synchrotron radiation model}
\label{sec:app}

We approximate the synchrotron radiation power density, $P'$, in the comoving 
frame (primed quantities) with a smoothly joined power-law at the characteristic 
frequencies. This takes the form
\begin{equation}
        \label{eq:radpow_approx_fast}
        P'(\nu') = 
                P'_{{\rm max,f}}\left[\left(\frac{\nu'}{\nu_c'}\right)^{(1/3)(-\kappa_1)}
                        +\left(\frac{\nu'}{\nu_c'}\right)^{(-1/2)(-\kappa_1)}\right]^{-1/\kappa_1}
                        \left[1+\left(\frac{\nu'}{\nu_m'}\right)^{(-1/2+p/2)(\kappa_2)}\right]^{-1/\kappa_2}
\end{equation}
for the fast cooling regime, $\nu'_c < \nu'_m$, and
\begin{equation}
        \label{eq:radpow_approx_slow}
        P'(\nu') = 
                P'_{{\rm max,s}}\left[\left(\frac{\nu'}{\nu_m'}\right)^{(1/3)(-\kappa_3)}
                        +\left(\frac{\nu'}{\nu_m'}\right)^{(-(p-1)/2)(-\kappa_3)}\right]^{-1/\kappa_3}
                        \left[1+\left(\frac{\nu'}{\nu_c'}\right)^{(-(p-1)/2+p/2)(\kappa_4)}\right]^{-1/\kappa_4}
\end{equation}
for the slow cooling regime, $\nu'_m < \nu'_c$.  Here $p$ is the electron
energy distribution index and $\nu'_m$ and $\nu'_c$
are the characteristic synchrotron frequencies corresponding,
respectively, to the
minimum Lorentz factor of the distribution,
\begin{equation}
        \gamma_m = \frac{p-2}{p-1}
        \left( \epsilon_e\frac{m_p}{m_e}(\Gamma - 1) + 1 \right)
\end{equation}
and the Lorentz factor above
which electrons are radiative, 
\begin{equation}
        \gamma_c = \frac{6\pi m_e c}{\sigma_T \Gamma(1+\beta) {B'}^2 t}.
\label{eq:gamma_c}
\end{equation}
Here, $m_p$ and $m_e$ are the mass of the proton and the electron,
respectively, $\epsilon_e$ is the fraction of thermal energy in the
electrons compared to the total thermal energy of the fireball,
$\sigma_T$ is the Thompson scattering cross section. Furthermore, $\beta$ 
is the dimensionless speed of the shock front and $B' = \sqrt{\epsilon_B 8\pi
(\Gamma - 1) n'}$ is the magnetic field strength,
with $\epsilon_B$ the fraction of magnetic field energy compared to
the total thermal energy of the fireball and $n' = 4\Gamma n$ is the
density in the comoving frame. The characteristic frequencies are given by
\begin{equation}
        \nu'_{i} = \chi_p \frac{3 \gamma_{i}^2 B'}{4 \pi m_e c},
\label{eq:nu_i}
\end{equation}
where $i$ is either $m$ or $c$.
The maximum values for the power density are given by
\begin{eqnarray}
        P'_{{\rm max,f}} &= &\phi_p \frac{2.234 e^3 n'B'}{m_e c^2}\\
        P'_{{\rm max,s}} &= &\phi_p \frac{11.17(p-1)}{3p-1}\frac{e^3 n'B'}{m_e c^2}
\end{eqnarray}
where, $e$, is the elementary charge and the parameters, $\phi_p$ and $\chi_p$, 
are introduced here as in \citet{WijersGalama1999} to account for an isotropic distribution 
of angles between the electron velocity and the magnetic field.  Comparing 
the results from these approximations with the more accurate calculations 
using the approach from \citet{Wuetal2004}, we find the following $p$-dependence 
of the exponents and coefficients in the fast cooling regime to be
\begin{eqnarray}
        \kappa_1 &= &2.37-0.3p\\
        \kappa_2 &= &14.7-8.68p+1.4p^2\\
        \phi_p &= &1.89-0.935p+0.17p^2\\
        \chi_p &= &0.06+0.28p.
\end{eqnarray}
In the slow cooling regime, we find similarly,
\begin{eqnarray}
        \kappa_3 &= &6.94-3.844p+0.62p^2\\
        \kappa_4 &= &3.5-0.2p\\
        \phi_p &= &0.54+0.08p\\
        \chi_p &= &0.455+0.08p.
\end{eqnarray}
The difference between these approximations and the more accurate
calculations is generally less than 10\% unless $\nu_c$ is close to
$\nu_m$. Then the difference can be up to 40\%. However, for most parameter
sets the characteristic frequencies are close, only very early
in the afterglow evolution and this is therefore
not a concern in the later evolution.  Even then, the difference
is mostly smoothed out because of the integration over the equal
arrival time surface. We find the difference between the two solutions
to be less than 10\% for most realistic cases.  We also remind the
reader that the electron energy distribution is an approximation and 
little is known about the detailed microphysics of the shocked
material.

\end{document}